\documentclass[11pt]{amsart}

\newtheorem{Theorem}{Theorem}[section]

\newtheorem{Lemma}[Theorem]{Lemma}

\newtheorem{Corollary}[Theorem]{Corollary}

\usepackage{hyperref}
\usepackage{stmaryrd}

\def\CC{{\mathbb C}}
\def\HH{{\mathbb H}}

\def\NN{{\mathbb N}}

\def\RR{{\mathbb{R}}}

\def\TT{{\mathbb T}}
\def\UU{{\mathbb U}}

\def\ZZ{{\mathbb Z}}
\def\vci{\vrule  width.02em height1.47ex depth-.0ex}
\def\11{{\rm\k{.2}\vci\k{-.37}1}}

\def\fin{{\begin{flushright}
\it{Q.E.D.}
\end{flushright}}}

\begin{document}

\title{The Dirac Sea for the Non-Separable Hilbert Spaces}
\address{Universit\'e de Bordeaux, Institut de Math\'ematiques, UMR CNRS 5251, F-33405 Talence Cedex}

\email{alain.bachelot@math.cnrs.fr}

\author{Alain BACHELOT}
\begin{abstract}
  We give a rigorous construction of the Dirac Sea for the fermionic
  quantization in the non-separable Hilbert spaces. These
  CAR-representations depend on the Axiom of Choice, hence are not
  unique, nevertheless they are unitarily equivalent to the classic
  Fock representation.
  \end{abstract}

\maketitle



  \section{Introduction}

  The old hole theory of Dirac, where the vacuum is replaced by a
  {\it sea} filled of all the negative energy states, continues to
  arouse the interest of physicists and mathematicians despite the
  severe criticism by Weinberg. For instance the Dirac sea plays an
  important role in the theory of the fermionic projector of Finster
  \cite{finster}, and recently it appears also in quantum
  cosmology \cite{berezin}. Of the mathematical point of view, a rigorous definition
  of such a sea is an interesting issue. In the case of the usual
  framework of the separable Hilbert spaces, an elegant construction is
  provided by a semi-infinite wedge product in the work of Dimock
  \cite{dimocksea} for the free Dirac equation, and for the external
  field problem by Deckert and co-authors \cite{deckert}. At the first
  glance, this approach heavely depends of the countability of the
  Hilbert basis and its generalization to the {\it non}-separable Hilbert
  spaces is not obvious. The purpose of this paper consists in showing
  that this goal is achievable with a little work of set theory. This one
  consists in defining in a coherent way a suitable notion describing the ``parity''
  of the size of all the subsets of an {\it infinite} set $X$. In
  short we prove the existence of a homomorphism $\pi$ from the
  Abelian group of the power set of $X$ endowed
  with the symetric difference, to $\ZZ/2\ZZ$, such that
  $\pi(A)=0$ (resp. $1$) if $A$ is a finite part of $X$ with an even
  (resp. odd) cardinal.  The proof that is based on the technics of
the ultrapowers, is strongly inspired by the theory of the numerosities of Benci {\it et
  alii} \cite{benciadv}, \cite{benciarist} (a shorter proof uses
powerful arguments from the Boolean Algebras Theory). As regards the axiomatic
framework, the price to pay to be able to treat the case of the non-separable
  Hilbert spaces, is a triple recourse to some consequences of the
  Axiom of Choice (see {\it e.g} \cite{howard}): 1) for the existence of a Hilbert
basis $(e_x)_{x\in X}$ on a non-separable Hilbert space; 2) for the existence of a linear order on the set
$X$ that indexes this basis; 3) in the proofs of the existence of $\pi$, we use one
of the following three consequences of Zorn's Lemma: {\it i)} the existence of a suitable
ultrafilter on the set of the finite parts of
$X$, {\it ii)}  Sikorski's extension theorem, {\it iii)} the
equivalence between ``completeness'' and ``injectivity'' for the
Boolean Algebras. As a consequence, our construction leads to a lot of different
quantizations. Fortunately, they are all unitarily equivalent to the
classic Fock quantization. Concerning the role of the non-separable
Hilbert spaces in Physics, the negative opinion of Streater and
Wightman in {\it PCT, Spin and Statistics, and All That}, is well
known. Nevertheless this issue is always matter to debate, see Earman
\cite{earman} for a highlighting discussion. We also remark that the
non-separable Hilbert spaces naturally arise in loop quantum gravity
(see {\it e.g.} \cite{bahr}).
\\

 We now introduce our strategy. We first fix the notations by  recalling the well known basics of the fermionic
   quantization (see {\it e.g.} \cite{dimock}, \cite{plymen}, \cite{thaller}). We consider a complex Hilbert
   space $(\mathfrak{h},<;>_{\mathfrak h})$  where $<;>_{\mathfrak h}$ is the inner product linear
   with respect to the second argument, and  we look for a Hilbert space $\mathfrak{H}$ and
   an antilinear map $\Psi$ from $\mathfrak{h}$ to the
space of the linear maps on $\mathfrak{H}$
satisfying the canonical
anticommutation relations (CAR): for any $u,v\in\mathfrak{h}$ we have
\begin{equation}
 \label{car}
 \left\{\Psi(u),[\Psi(v)]^*\right\}=<u;v>_{\mathfrak{h}}Id_{\mathfrak{H}},
\end{equation}
 \begin{equation}
 \label{carde}
 \left\{\Psi(u),\Psi(v)\right\}=0,
\end{equation}
where $\{A,B\}:=AB+BA$ is the
anticommutator of two operators $A$ and $B$. Taking the adjoint of (\ref{carde}) we also have
\begin{equation}
 \label{cardestar}
 \left\{[\Psi(u)]^*,[\Psi(v)]^*\right\}=0.
\end{equation}
Another important consequence of the CAR is that $\Psi(u)$ belongs to
the space $\mathcal{L}(\mathfrak{H})$ of the bounded linear maps on
$\mathfrak{H}$ and we have
\begin{equation}
 \label{}
 \Vert\Psi(u)\Vert_{\mathcal{L}(\mathfrak{H})}=\Vert[\Psi(u)]^*\Vert_{\mathcal{L}(\mathfrak{H})}=\Vert
 u\Vert_{\mathfrak h}.
\end{equation}

The classic representation of the CAR on $\mathfrak{h}$ is given by choosing the
antisymetric Fock space
\begin{equation}
 \label{ffs}
 \mathfrak{H}=\mathcal{F}^{\wedge}(\mathfrak{h}):=\bigoplus_{n=0}^{\infty}\mathfrak{h}^{\wedge n},
\end{equation}
and $\Psi=a$, the usual annihilation operator that is the adjoint of the
creation operator  $a^*$, which is nicely expressed by using the wedge product:
\begin{equation}
 \label{fpsi}
a^*(u)\left(v_1\wedge v_2\wedge...\wedge
   v_n\right)=u\wedge v_1\wedge v_2\wedge...\wedge v_n.
\end{equation}
Up to an unitary transform, the Fock quantization is the unique
irreductible representation of the CAR on $\mathfrak{h}$.\\

We are now ready to introduce the issue of the Dirac sea in this
framework. The idea is that, unlike the classic Dirac quantum field,
which annihilates a particle but {\it creates} an anti-particle, the
fermionic quantum field should just  be an
annihilation operator in a suitable sense: the creation of an
antiparticle  in the Fock quantization should
be understood as the creation of a hole in the Dirac sea, {\it i.e.} the
annihilation of a state of negative energy, the Dirac sea being filled
with all these states. This idea has been
rigorously implemented by  Dimock in \cite{dimocksea} when
$\mathfrak{h}=\mathfrak{h}_-\oplus \mathfrak{h}_+$ is {\it separable}. In
order to point out the role of the separability, we
briefly describe his approach based on the semi-infinite wedge products. In the sequel, if $X$ is a set of
integers or
ordinals, we put $X^*:=X\setminus\{0\}$. Given a Hilbert basis
$(e_{\pm j})_{j\in\NN^*}$ of $\mathfrak{h}_{\pm}$, we consider the Hilbert completion
$\mathfrak{H}$ of the
free vector space
spanned by the formal symbols
\begin{equation}
 \label{eI}
 e_I:=e_{i_1}\wedge e_{i_2}\wedge...
\end{equation}
where $I=\{i_s,\;\;s\in\NN^*\}$ and
$(i_s)_{s\in\NN^*}\in(\ZZ^*)^{\NN^*}$ is strictly decreasing sequence
(for the usual order on $\ZZ$)
satisfying for $s$ large enough
\begin{equation}
 \label{ss+1}
 i_{s+1}=i_s-1.
\end{equation}
Then the Dirac sea is
the vector
\begin{equation}
 \label{}
 \Omega_D:=e_{-1}\wedge e_{-2}\wedge e_{-3}\wedge...
\end{equation}
and the quantum field is the antilinear map $\Psi$ from $\mathfrak{h}$ to $\mathcal{L}(\mathfrak{H})$
defined by:
\begin{equation}
 \label{psidi}
 \Psi(e_j)e_{i_1}\wedge e_{i_2}\wedge e_{i_3}\wedge...=\left\{
   \begin{array}{c}
{0}\; \;if\;\; \forall s\in\NN, \;\;j\neq i_s,\\
(-1)^{s+1} e_{i_1}\wedge e_{i_2}\wedge... \wedge e_{i_{s-1}}\wedge
     e_{i_{s+1}}\wedge...\;\;if\;\;\exists s\in\NN,\;\;j=i_s.
   \end{array}
   \right.
\end{equation}
The key point in definition (\ref{psidi}) is the factor $(-1)^s$ that
is well defined thanks to the obvious fact that $s=s(j,I)$ defined by
\begin{equation}
 \label{sij}
 j=i_s
\end{equation}
is a {\it finite} ordinal that is just the cardinal of the subset
\begin{equation}
 \label{}
X(j,I):=\{\;i\in I,\;\;i\geq j\}.
\end{equation}
A crucial property to obtain the CAR is that
\begin{equation}
 \label{moinzun}
k>j,\;k\in I\Rightarrow (-1)^{ s(j,I)}=-(-1)^{s(j,I\setminus\{k\})},
\end{equation}
since
\begin{equation}
 \label{skkun}
 k>j,\;k\in I\Rightarrow s(j,I)=s(j,I\setminus\{k\})+1.
\end{equation}
\\

The situation drastically changes when
$\mathfrak{h}$ is a  {\it non}-separable Hilbert space. In this case we
have to consider a Hilbert basis $(e_j)_{j\in X}$ of $\mathfrak{h}$, where $X$ is a
totally ordered set of uncountable cardinal $\mid
X\mid=\aleph>\aleph_0$, and, instead of the {\it countable} wedge products, the straight generalization of (\ref{eI})
and (\ref{ss+1}) would consist in considering the formal symbols
\begin{equation}
 \label{}
 e_I=\bigwedge_{j\in I}e_j
\end{equation}
where $I=\{i_s,\;\;s\in\aleph^*\}$ is {\it uncountable}, and $(i_s)_{s\in\aleph^*}\in
X^{\aleph^*}$ is a strictly decreasing generalized sequence
satisfying for some $a\in X$
\begin{equation}
 \label{a}
x<a\Rightarrow x\in I.
\end{equation}
Now given $I$ and $j\in X$, $s=s(j,I)$ solution of (\ref{sij})  is an ordinal that belongs to $\aleph$ and we have to define
$(-1)^s$ satisfying (\ref{moinzun}). Of course the difficulty arises
when $s$ is an {\it infinite} ordinal. We could try to use the Cantor's normal
form theorem that assures that $s$ can be uniquely expressed as
$$
s=\lambda+N,
$$
with $N\in\NN$ and $\lambda$ is a limit ordinal. Deciding classically
that the limit ordinals are even, we could define
\begin{equation}
 \label{defw}
 (-1)^s:=(-1)^N.
\end{equation}
Unfortunately this definition does not assure the key point
(\ref{moinzun}) since (\ref{skkun}) can be wrong due to the absorbing
property of the ordinal addition
$$
1+\lambda=\lambda.
$$
For an elementary example, we can take $I=\{-n;\;\;n\in\NN^*\}\cup
\{-\omega\}$ where $-\omega<-n$ for any integer $n$. Then with
$j=-\omega$ and $k=-n$ for some integer $n$, we have $s(-\omega,I)=\aleph_0$ and also
 $s(-\omega,I\setminus\{-n\})=\aleph_0$. Therefore with definition
 (\ref{defw}) we have
 $$
(-1)^{s(-\omega,I)}=1=(-1)^{s(-\omega,I\setminus\{-n\})}
$$
that contradicts (\ref{moinzun}). We conclude that the ordinal
calculus is not sufficient to associate $1$ or $-1$ to the set
$X(j,I)$ in such a way that (\ref{moinzun}) is satisfied even if
$X(j,I)$ is infinite. To overcome
this difficulty, we simply remark that in the separable case for which
$X(j,I)$ is finite,
$(-1)^{s(j,I)}=1$ if the cardinal of $X(j,I)$ is even and
$(-1)^{s(j,I)}=-1$ if the cardinal of $X(j,I)$ is odd. Finally we are
led to ask a somewhat weird question: what is the parity of the size
of an infinite set? Clearly, in the infinite case, the cardinal is not a tool
sufficiently subtle to distinguish the size of $X(j,I)$ from the size
of $X(j,I\setminus\{k\})$. In fact a refined concept of size of an
infinite set has been introduced by Benci and co-authors \cite{benciadv},
\cite{benciarist}. It is a hypernatural $s$ and we could define $(-1)^s$
in the framework of this theory of the numerosities, from the parity of $s$, but we prefer
to give a direct construction in the next section. We introduce the
quantum fields in an abstract setting in the third part. We present
the application to the Dirac theory in the last section.


  \section{Parity of an infinite set}
   We first introduce some notations. The cardinal of a set $X$ is denoted $\mid
  X\mid$, $\mathcal{P}(X)$ is its power set, and $\mathcal{P}_F(X)$ is
  the set of the finite parts of $X$. The symetric difference $\Delta$
  is defined by
  $$
A,B\in\mathcal{P}(X),\;\;A\Delta B:=(A\cup B)\setminus(A\cap B).
  $$


Given an infinite set $X$ we look for a homomorphism $\pi$ from
$(\mathcal{P}(X),\Delta)$ to $(\ZZ/2\ZZ,+)$ such that
\begin{equation}
 \label{singleton}
 \forall a\in X,\;\;\pi(\{a\})=1.
\end{equation}
The requirement
\begin{equation}
 \label{}
 \pi(A\Delta B)=\pi(A)+\pi(B)
\end{equation}
is equivalent to the couple of properties
\begin{equation}
 \label{piplus}
 \forall A,B\in \mathcal{P}(X),\;\;A\cap
 B=\emptyset\Rightarrow\pi\left(A\cup B\right)=\pi(A)+\pi(B)\;\;in\;\;\ZZ/2\ZZ,
\end{equation}
\begin{equation}
 \label{pimoins}
 \forall A,B\in \mathcal{P}(X),\;\;B\subset
 A\Rightarrow\pi\left(A\setminus B\right)=\pi(A)-\pi(B)\;\;in\;\;\ZZ/2\ZZ.
\end{equation}
If $\pi$ exists, then (\ref{singleton}) and (\ref{piplus}) imply that
for $A\in\mathcal{P}_F(X)$, 
$\pi(A)$ is the usual parity of the cardinal of $A$, defined as $0$ if it is
even and $1$ if it is odd. The extension of $\pi$ from
$\mathcal{P}_F(X)$ to the whole $\mathcal{P}(X)$ is not at all obvious: its existence
depends on the Axiom of Choice, and then such a $\pi$ is not unique
since we may impose $\pi(X)=0$ or $\pi(X)=1$ as well.

\begin{Theorem}[Parity function]
 We consider an infinite set $X$ and $p\in\ZZ/2\ZZ$. Then there exists a homomorphism
 $\pi:(\mathcal{P}(X),\Delta)\rightarrow (\ZZ/2\ZZ,+)$ satisfying
 (\ref{singleton}) and
 \begin{equation}
 \label{pip}
 \pi(X)=p.
\end{equation}
 \label{parity}
\end{Theorem}

We present two proofs of this result. The first one, rather
long and pedestrian, uses a suitable Ultrafilter. The second one,
short and elegant, has been suggested by the anonymous referee; it is
based on a powerful tool of the Boolean Algebras Theory: either
Sikorski's extension
theorem, or the injectivity of $\{0,1\}$.\\

{\it First proof.} We identify the additive group $\ZZ/2\ZZ$ and $\{0,1\}$ and we denote $\mathcal{I}:=\mathcal{P}_F(X)\setminus\{\emptyset\}$. Given
$p\in\{0,1\}$, we define for any $i\in\mathcal{I}$
\begin{equation}
 \label{}
 C_p(i):=\{j\in\mathcal{I};\;\;i\subset j,\;\;\mid j\mid\in 2\NN+p\}\in\mathcal{P}(\mathcal{I})\setminus\{\emptyset\}.
\end{equation}
We have for any $i,j\in\mathcal{I}$
\begin{equation}
 \label{}
 C_p(i)\cap C_p(j)=C_p(i\cup j),
\end{equation}
hence the family $\mathcal{B}_p:=(C_p(i))_{i\in\mathcal{I}}$ is a
filter basis on $\mathcal{I}$. Applying the Axiom of
Choice, we consider an ultrafilter $\mathcal{U}_p$ containing
$\mathcal{B}_p$. $\{0,1\}^{\mathcal{I}}$ being the additive group of the maps
from $\mathcal{I}$ to $\ZZ/2\ZZ$, we consider its ultrapower
\begin{equation}
 \label{}
 \{0,1\}^{\mathcal{I}}/{\mathcal{U}_p}:=\{\dot{\varphi},\;\;\varphi\in  \{0,1\}^{\mathcal{I}}\},\;\;\dot{\varphi}:=\{\psi\in  \{0,1\}^{\mathcal{I}};\;\;\{i;\;\;\varphi(i)=\psi(i)\}\in\mathcal{U}_p\}.
\end{equation}
In fact this ultrapower is just a pair set: since $\mathcal{U}_p$ is
an ultrafilter, either $\{i\in \mathcal{I}; \;\;\varphi(i)=0\}$ or
$\{i\in \mathcal{I}; \;\;\varphi(i)=1\}$ belongs to
$\mathcal{U}_p$. We denote $p_{\varphi}$ the element of $\{0,1\}$ such
that $\{i\in \mathcal{I}; \;\;\varphi(i)=p_{\varphi}\}\in
\mathcal{U}_p$. Now for $\psi\in\dot{\varphi}$ we have
$$
 \{i\in\mathcal{I};\;\;\psi(i)=p_{\varphi}\}\supset \{i\in\mathcal{I};\;\;\psi(i)=\varphi(i)\}\cap\{i\in\mathcal{I};\;\;\varphi(i)=p_{\varphi}\}\in\mathcal{U}_p,
 $$
 hence $\{i\in\mathcal{I};\;\;\psi(i)=p_{\varphi}\}\in\mathcal{U}_p$ and thus $p_{\varphi}=p_{\psi}$, and we may introduce the map $p$
 \begin{equation}
 \label{}
 p:\dot{\varphi}\in \{0,1\}^{\mathcal{I}}/{\mathcal{U}_p}\longmapsto p(\dot{\varphi}):=p_{\varphi}\in\{0,1\}.
\end{equation}
$p$ is a group isomorphism. It is obviously surjective, it is a
homomorphism since
$$
\{i;\;\;\varphi(i)+\psi(i)=p_{\varphi}+p_{\psi}\}\supset\{i;\;\;\varphi(i)=p_{\varphi}\}\cap\{i;\;\;\psi(i)=p_{\psi}\}\in\mathcal{U}_p,
$$
hence $\{i;\;\;\varphi(i)+\psi(i)=p_{\varphi}+p_{\psi}\}\in\mathcal{U}_p$ and thus $p(\dot{\varphi}+\dot{\psi})=p(\dot{\varphi})+ p(\dot{\psi})$, and finally it is injective
since
$$
p(\dot{\varphi})=0\Leftrightarrow p_{\varphi}=0\Leftrightarrow\{i;\;\;\varphi(i)=0\}\in\mathcal{U}_p\Leftrightarrow\dot{\varphi}=\dot{0}.
$$

To any $A\in\mathcal{P}(X)$ we associate
$\varphi_A\in\{0,1\}^{\mathcal{I}}$ by
$$
\forall i\in\mathcal{I},\;\;\varphi_A(i)=0\Leftrightarrow\mid A\cap i\mid\in 2\NN,\;\; \varphi_A(i)=1\Leftrightarrow\mid A\cap i\mid\in 2\NN+1,
$$
and we define
\begin{equation}
 \label{}
 \pi(A):=p(\dot{\varphi}_A).
\end{equation}
We remark that for any $a\in X$, we have
\begin{equation}
 \label{}
 \{i\in\mathcal{I};\;\;a\in i\}\in\mathcal{U}_p
\end{equation}
since
$$
\{i\in\mathcal{I};\;\;a\in i\}\supset C_p(\{a\})\in\mathcal{U}_p.
$$
We have
$$
\varphi_{\{a\}}(i)=1\Leftrightarrow i\in\{j\in\mathcal{I};\;\;a\in j\},
$$
hence $p_{\varphi_{\{a\}}}=1$ and we deduce that (\ref{singleton}) is
satisfied.
Now for $A,B\in\mathcal{P}(X)$ with $A\cap B=\emptyset$, we have
$$
\{i;\;\varphi_{A\cup B}(i)=p_{\varphi_A}+p_{\varphi_B}\}\supset
\{i;\;\varphi(A)(i)=p_{\varphi_A}\}\cap \{i;\;\varphi(B)(i)=p_{\varphi_B}\}\in\mathcal{U}_p
$$
hence $p_{\varphi_{A\cup B}}=p_{\varphi_A}+p_{\varphi_B}$ and (\ref{piplus}) is
  established. Moreover (\ref{pimoins}) is a consequence of
  (\ref{singleton}) and (\ref{piplus}) since if $B\subset A$ we have
  $\pi(A)=\pi(A\setminus B)+\pi(B)$. Finally we have
  $$
\{i;\;\varphi_X(i)=p\}=\{i;\;\mid i\mid\in 2\NN+p\}\supset C_p(\{a\})\in\mathcal{U}_p
$$
hence $\{i;\;\varphi_X(i)=p\}\in\mathcal{U}_p$, and thus $\pi(X)=p$.
\fin

{\it Second proof.}
We consider the collection $\mathcal{A}$ of the subsets of $X$ that are either finite or co-finite:
$\mathcal{A}$ is a Boolean algebra. We define a map
$
\pi_0:\mathcal{A}\rightarrow \{0,1\}
$
as follows:  If $A$ is a finite subset of $X$, then $\pi_0(A)=0$ if $\mid A\mid$ is even, otherwise $\pi_0(A)=1.$ If $A$ is a co-finite subset of X, then $\pi_0(A)=p$ if $\mid X\setminus A\mid$ is even, otherwise $\pi_0(A)=1-p. $
 $\pi_0$ is clearly a Boolean homomorphism from $\mathcal{A} $ to $\{0,1\}$
considered as a Boolean Algebra. Since $\{0,1\}$ is trivially a complete
algebra, $\pi_0$ can be extended to a Boolean homomorphism $\pi$ from
$\mathcal{P}(X)$ to $\{0,1\}.$ The existence of $\pi$ is assured
either by Sikorski's extension theorem (\cite{sikorski}, p.141,
Theorem 33.1) or by the injectivity of
$\{0,1\}$ due to the theorem on the equivalence between the completeness
and the injectivity (\cite{halmos}, p.141, Theorem 19).

\fin


We now are ready to deduce the main tool for the quantization, the
``$\epsilon$-functions''  that replace $(-1)^s$ in the definition of
the quantum fields.

\begin{Corollary}[``$\epsilon$-function'']
  Given a partially ordered set $(X,\leq)$, there exists a map
 \begin{equation}
 \label{}
 \epsilon:\;I\in\mathcal{P}(X)\longmapsto \epsilon_I\in \{-1,+1\}^I
\end{equation}
satisfying
\begin{equation}
 \label{epsing}
 \forall x\in X,\;\;\epsilon_{\{x\}}(x)=1,
\end{equation}
\begin{equation}
 \label{jkmoins}
 j,k\in I,\;j<k\Rightarrow \epsilon_I(k)=-\epsilon_{I\setminus\{j\}}(k),
\end{equation}
\label{}
\begin{equation}
 \label{jkplus}
 j\in I,\;k\in X,\;\;j<k\Rightarrow \epsilon_I(j)=\epsilon_{I\cup\{k\}}(j).
\end{equation}
 \label{corolpa}
\end{Corollary}

{\it Proof.} We choose a parity function $\pi$  given by the previous
theorem. For $I\in\mathcal{P}(X)$, $x\in X$, we put
$I_x:=I\cap\{y\in X;\;\;y\leq x\}$. We obviously have: for any $x\in
X$, 
$$
\{x\}_x=\{x\},\;\;\pi(\{x\}_x)=1,
$$
for any $j,k\in I$ with $j<k$,
$$
I_k=(I\setminus\{j\})_k\cup\{j\},\;\;\pi(I_k)=\pi((I\setminus\{j\})_k)+1,
$$
for any $j\in I$ and $k>j$
$$
I_j=(I\cup\{k\})_j.
$$
Therefore it is sufficient to define for $I\in\mathcal{P}(X)$, $j\in I$
\begin{equation*}
 \label{}
 \epsilon_I(j):=(-1)^{\pi(I_j)+1}.
\end{equation*}

\fin

To make the link with the countable case, we consider $X=\ZZ^*$
 and $I=\{i_s,\;\;s\in\NN^*\}$ where
$(i_s)_{s\in\NN^*}\in(\ZZ^*)^{\NN^*}$ is a strictly decreasing
sequence (for the usual order on $\ZZ$)
satisfying (\ref{ss+1})  for $s$ large enough. We now endow $\ZZ^*$ with
the {\it reverse} order of the usual order, then $\epsilon_I(j):=(-1)^{s+1}$ where
$j=i_s$, satisfies (\ref{epsing}), (\ref{jkmoins}) and (\ref{jkplus}).

\section{Abstract quantization}

We now consider an infinite ordered set $(X,\leq)$ and a subset
$\mathcal{I}\subset\mathcal{P}(X)$ such that
\begin{equation}
 \label{}
 X\;is \;totally\;ordered,
\end{equation}
\begin{equation}
 \label{III}
 \forall I\in\mathcal{I},\;\;\forall A\in\mathcal{P}_F(X),\;\;I\cup A\in\mathcal{I},\;\; I\setminus A\in\mathcal{I}.
\end{equation}
We also take an $\epsilon$-function given by Corollary \ref{corolpa}.\\

For the quantization process, $X$ enumerates a Hilbert basis
$(e_x)_{x\in X}$ of the Hilbert space $\mathfrak{h}$ of the classical
fields, and $\mathcal{I}$ indexes a Hilbert basis of the Hilbert space
$\mathfrak{H}$ on which the quantum
fields act. The existence of a linear order on $X$ is assured by the
Ordering Principle which is strictly weaker than the Zermelo
Axiom. The $\epsilon$-function will play the role of $(-1)^s$ in the
definition of the quantum fields.\\

We introduce the equivalence relation
$\mathcal{R}$ on the abelian group $(\mathcal{P}(X),\Delta)$
associated to the subgroup $\mathcal{P}_F(X)$. $\mathcal{R}$ is defined
by
\begin{equation}
 \label{}
 \forall A,B\in\mathcal{P}(X),\;\;A\mathcal{R}B\Leftrightarrow A\Delta B\in\mathcal{P}_F(X).
\end{equation}
Since $A=(A\cap B)\sqcup\left(A\cap(A\Delta B)\right)$ we also have
\begin{equation}
 \label{}
 A {\mathcal R}B\Leftrightarrow\exists A',B',\in\mathcal{P}_F(X),\;\;C\in
 \mathcal{P}(X),\;
 A=A'\sqcup C,\;\;B=B'\sqcup C.
\end{equation}
${\mathcal R}$ is an equivalence relation on
$\mathcal{P}(X)$ for which we denote ${\mathcal
  P}(X )/\mathcal{P}_F(X)$ its quotient
set and $[A]_X$ the equivalence class of $A\in\mathcal{P}(X)$.
The equivalence class $[A]_X$ is also described as
\begin{equation}
 \label{}
 I\in[A]_X\Leftrightarrow \exists I_F\in\mathcal{P}_F(X),\;I=A\Delta I_F,
\end{equation}
and the map 
\begin{equation}
 \label{}
 I_F\in\mathcal{P}_F(X)\longmapsto I=A\Delta I_F\in[A]_X
\end{equation}
is a bijection from $\mathcal{P}_F(X)$ onto $[A]_X$.
We note that $[\emptyset]_X=\mathcal{P}_F(X)$ and
$[X]_X$ is just the set of the co-finite subsets of
$X$.
Putting $I_-:=I_F\cap A$ and $I_+:=I_F\cap(X\setminus A)$, we also have
\begin{equation}
 \label{decomposI}
 \forall A\in\mathcal{P}(X),\;\;\forall
   I\in[A]_X,\;\;\exists I_-\in\mathcal{P}_F(A),\;\; \exists
   I_+\in\mathcal{P}_F(X\setminus A),\;\;I=(A\setminus I_-)\sqcup I_+.
\end{equation}
We obviously have for any $A\subset X$:
\begin{equation}
 \label{}
 \mid[A]_X\mid=\mid X\mid,
\end{equation}
and since
$$
\mathcal{P}(X)=\bigsqcup_{[A]_X\in\mathcal{P}(X)/\mathcal{P}_F(X)}[A]_X,\;\;\mid
\mathcal{P}(X)\mid=2^{\mid X\mid},
$$
we have
\begin{equation}
 \label{}
 \mid \mathcal{P}(X)/\mathcal{P}_F(X)\mid=2^{\mid X\mid}.
\end{equation}
Moreover $\mathcal{I}\subset\mathcal{P}(X)$ satisfies (\ref{III}) iff
\begin{equation}
 \label{}
 \mathcal{I}=\bigcup_{A\in\mathcal{I}}[A]_X.
\end{equation}

Now we introduce the free vector space $\mathcal{V(I)}$ spanned by
$\mathcal{I}$, {\it i.e.} we consider the elements $I$ of
$\mathcal{I}$ as vectors denoted $e_I$, and  $\mathcal{V(I)}$
becomes a pre-Hilbert space if we decide $\mathcal{I}$ is an orthonormal
basis.\\

The fundamental example arising for the quantization in the separable
case is given by:
\begin{equation}
 \label{}
 X=\ZZ^*=\ZZ^-\sqcup\ZZ^+,\;\;\ZZ^{\pm}:=\{n\in\ZZ;\;\;\pm n\geq 1\},
\end{equation}
and $<$ is just
the {\it reverse} of the usual strict order on $\ZZ$.
We choose $\mathcal{I}$ to be the set of the ``maya diagrams''
(see {\it e.g.} \cite{foda}):
\begin{equation}
 \label{exi}
 \mathcal{I}=\{I=(\ZZ^-\setminus I^-)\sqcup I^+;\;\;I^{\pm}\in\mathcal{P}_F(\ZZ^{\pm})\}=[\ZZ^-]_{\ZZ^*}.
\end{equation}
The {\it Dirac sea} is the vector
\begin{equation}
 \label{didis}
 \Omega_D:=e_{\ZZ^-}\in\mathcal{V}([\ZZ^-]_{\ZZ^*}).
\end{equation}

We easily generalize this example to take into account the
non-separable Hilbert spaces for which $X$ is uncountable. Given two
infinite ordinals $\Lambda_{\pm}$ we introduce
\begin{equation}
 \label{}
 X=\Lambda_-^*\sqcup\Lambda_+^*
\end{equation}
endowed with the total strict order $<$ defined by
\begin{equation}
 \label{}
 \forall \lambda_{\pm},\lambda'_{\pm}\in\Lambda_{\pm}^*,\;\;\lambda_+<\lambda_-,\;\;\lambda_-<\lambda'_-\Leftrightarrow\lambda_-\prec\lambda'_-,\;\; \lambda_+<\lambda'_+\Leftrightarrow\lambda'_+\prec\lambda_+,
\end{equation}
where $\prec$ is the usual strict well order on the ordinals. To
generalize (\ref{exi}) and (\ref{didis}), we can take
\begin{equation}
 \label{exig}
 \mathcal{I}=[\Lambda_-^*]_{\Lambda_-^*\sqcup\Lambda_+^*}=\{I:(\Lambda_-^*\setminus
 I_-)\sqcup I_+;\;\;I_{\pm}\in{\mathcal P}_F(\Lambda_{\pm}^*)\},
\end{equation}
\begin{equation}
 \label{didins}
 \Omega_D:=e_{\Lambda_-^*}\in\mathcal{V}([\Lambda^*_-]_{\Lambda_-^*\sqcup\Lambda_+^*}).
\end{equation}

We now choose the Hilbert framework. Given a set $E$, we introduce
the Hilbert space of the complex-valued square integrable functions on
$E$ with respect to the counting measure, that is also the Hilbert closure of the
free vector space $\mathcal{V}(E)$  spanned by $E$,
\begin{equation}
 \label{}
 l^2(E):=\left\{u\in\CC^E;\;\;\Vert u\Vert:=\left(\sum_{x\in E}\mid u(x)\mid^2\right)^{\frac{1}{2}}<\infty\right\}.
\end{equation}
We use the canonical Hilbert basis
\begin{equation}
 \label{}
 \mathcal{B}(E):=\left\{e_x\in\CC^E;\;x\in E\right\},\;\forall x,
 y\in E,\;\;e_x(y)=\delta_{x,y},
\end{equation}
here $\delta_{x,y}$ is the Kronecker symbol ($\delta_{x,x}=1$,
$\delta_{x,y}=0$ if $x\neq y$). Then $u\in l^2(E)$ can be written as
\begin{equation}
 \label{}
 u=\sum_{x\in E}c_xe_x,\;\;c_x\in\CC,
\end{equation}
where $\{x;\;c_x\neq 0\}$ is countable and
\begin{equation}
 \label{}
 \Vert u\Vert^2=\sum_{x\in E}\mid c_x\mid^2.
\end{equation}
For $F\subset E$ we denote $P_F$ the orthogonal projector on $l^2(F)$
\begin{equation}
 \label{}
 P_F(u)=\sum_{x\in F}<e_x;u>e_x.
\end{equation}
For any infinite set $X$ and $A\subset X$, we have
\begin{equation}
 \label{}
 l^2(X)=l^2(A)\oplus l^2(X\setminus A),\;\;l^2(\mathcal{P}(X))=\bigoplus_{[A]_X\in\mathcal{P}(X)/\mathcal{P}_F(X)}l^2([A]_X),
\end{equation}
and since the elements $I$ of $[A]_X$ can be indexed by $I_F\in \mathcal{P}_F(X)$ or
$(I_-,I_+)\in\mathcal{P}_F(A)\times\mathcal{P}_F(X\setminus A)$, we
can identify the following spaces with natural isometries:
\begin{equation*}
 \label{}
l^2([A]_X),\; l^2(\mathcal{P}_F(X)),\;l^2(\mathcal{P}_F(A)\times
\mathcal{P}_F(X\setminus A)),\;l^2(\mathcal{P}_F(A))\otimes
l^2(\mathcal{P}_F(X\setminus A)),\;l^2([\emptyset]_A)\otimes
l^2([\emptyset]_{X\setminus A}).
\end{equation*}


We now construct two maps, $\psi$, the ``annihilation
operator'',  and $\psi^*$, the ``creation operator'',  from
$l^2(X)$ to the space $\mathcal{L}(l^2(\mathcal{P}(X)))$ of the
bounded linear maps on $l^2(\mathcal{P}(X))$.
Given $j\in X$ and $I\in\mathcal{P}(X)$, we put:
\begin{equation}
 \label{psi}
 \psi(e_j)e_{I}:=\left\{
   \begin{array}{cc}
     0 & \text{if}\;\;j\notin I,\\
     \epsilon_I(j)e_{ I\setminus\{j\}}& \text{if} \;\;j\in I,
   \end{array}
   \right.
 \end{equation}

 \begin{equation}
 \label{psistar}
 \psi^*(e_j)e_{I}:=\left\{
   \begin{array}{cc}
     0 & \text{if}\;\;j\in I,\\
     \epsilon_{I\cup\{j\}}(j)e_{I\cup\{j\}}& \text{if} \;\;j\notin I.
   \end{array}
   \right.
 \end{equation}
Since $\psi(e_j)$ and $\psi^*(e_j)$ map $\mathcal{B}(\mathcal{P}(X))$ into $\mathcal{B}(\mathcal{P}(X))\cup\{0\}$,  we can extend
$\psi(e_j)$ and $\psi^*(e_j)$ as bounded linear operators on
$l^2(\mathcal{P}(X))$ with
\begin{equation}
 \label{}
 \Vert \psi(e_j)\Vert_{\mathcal{L}(l^2(\mathcal{I}))}=1,\;\;  \Vert \psi^*(e_j)\Vert_{\mathcal{L}(l^2(\mathcal{I}))}=1.
\end{equation}

We note that (\ref{epsing}) implies that
\begin{equation}
 \label{}
 \psi(e_j)e_{\{j\}}=e_{\emptyset},\;\;\psi^*(e_j)e_{\emptyset}=e_{\{j\}}.
\end{equation}
The physical interpretation of these operators is classic: $\psi(e_j)$
annihilates the state $e_{\{j\}}$ and $\psi^*(e_j)$
creates the state $e_{\{j\}}$.
We also have with (\ref{epsing}), (\ref{jkmoins}) and (\ref{jkplus})
\begin{equation}
 \label{}
 j<k\Rightarrow\psi^*(e_k)\psi^*(e_j)e_{\emptyset}=-e_{\{j,k\}},\;\;\psi^*(e_j)\psi^*(e_k)e_{\emptyset}=e_{\{j,k\}}
\end{equation}
More generally, we arrive to the fundamental algebraic  properties: $\psi(e_j)$ and
$\psi^*(e_j)$ are adjoint to each other and satisfy the canonical
anticommutation relations (CAR).

\begin{Lemma}
  For any $j,k\in X$, we have
  \begin{equation}
 \label{adjj}
\psi^*(e_j) =\left[\psi(e_j)\right]^*,
\end{equation}
\begin{equation}
 \label{car}
 \{\psi(e_j),\psi^*(e_k)\}=\delta_{jk}Id,
\end{equation}
\begin{equation}
 \label{carpsi}
 \{\psi(e_j),\psi(e_k)\}=0,
\end{equation}
\begin{equation}
 \label{carstar}
 \{\psi^*(e_j),\psi^*(e_k)\}=0.
\end{equation}
For any $A\subset X$, $\psi(e_j)$ and $\psi^*(e_k)$ leave invariant
$l^2([A]_X)$.
\end{Lemma}

{\it Proof.} To prove (\ref{adjj}) it is sufficient to establish for
any $I,J\in\mathcal{I}$ that:
 \begin{equation}
 \label{toto}
 \left\langle\psi(e_j) e_I; e_J\right\rangle=
 \left\langle e_I;  \psi^*(e_j)e_J\right\rangle.
\end{equation}
We have
\begin{equation}
 \label{}
\left\langle\psi(e_j) e_I; e_J\right\rangle=\left\{
   \begin{array}{cc}
     0 & \text{if}\;\;j\notin I,\\
     \epsilon_I(j)\delta_{ I\setminus\{j\},J}& \text{if} \;\;j\in I,
   \end{array}
   \right.
 \end{equation}
\begin{equation}
 \label{}
\left\langle e_I;  \psi^*(e_j)e_J\right\rangle=\left\{
   \begin{array}{cc}
     0 & \text{if}\;\;j\in J,\\
     \epsilon_{J\cup\{j\}}(j)\delta_{J\cup\{j\},I}& \text{if} \;\;j\notin J.
   \end{array}
   \right.
\end{equation}
Therefore $\left\langle\psi(e_j) e_I; e_J\right\rangle$ and
$\left\langle e_I;  \psi^*(e_j)e_J\right\rangle$ are non zero iff
$j\notin J$ and $I=J\cup\{j\}$. Moreover in this case they are equal
to  $\epsilon_I(j)$. That proves (\ref{toto}).\\

Now we consider $j\in I$  and we evaluate
$$
\psi(e_j)\psi^*(e_j)e_I=0,\;\; \psi^*(e_j)\psi(e_j)e_I=\epsilon_I(j)\psi^*(e_j)e_{I\setminus\{j\}}=\epsilon_I(j)\epsilon_I(j)e_{I}=e_I
$$
hence we have
\begin{equation}
 \label{ji}
 j\in I\Rightarrow \{\psi(e_j),\psi^*(e_j)\}e_I=e_I.
\end{equation}
Then we consider $j\notin I$ and we evaluate
$$
\psi^*(e_j)\psi(e_j)e_I=0,\;\; \psi(e_j)\psi^*(e_j)e_I=\epsilon_{I\cup\{j\}}(j)\psi(e_j)e_{I\cup\{j\}}=\epsilon_{I\cup\{j\}}(j)\epsilon_{I\cup\{j\}}(j)e_{I}=e_I
$$
that implies
\begin{equation}
 \label{jin}
 j\notin I\Rightarrow \{\psi(e_j),\psi^*(e_j)\}e_I=e_I.
\end{equation}

Now we take $j\neq k$. 1) We first assume $j\in I$ and $k\notin I$. We
compute:
\begin{equation}
 \label{jkI}
\psi(e_j)\psi^*(e_k)e_I=\epsilon_{I\cup\{k\}}(j)\epsilon_{I\cup\{k\}} (k)e_{(I\setminus\{j\})\cup\{k\}},
\end{equation}
\begin{equation}
 \label{jkI}
 \psi^*(e_k)\psi(e_j)e_I=\epsilon_I(j)\epsilon_{(I\setminus\{j\})\cup\{k\}}(k)e_{(I\setminus\{j\})\cup\{k\}}.
\end{equation}
If $j<k$, then $\epsilon_{I\cup\{k\}}(j)=\epsilon_I(j)$ by
(\ref{jkplus}),
$\epsilon_{(I\setminus\{j\})\cup\{k\}}(k)=-\epsilon_{I\cup\{k\}} (k)$
by (\ref{jkmoins}) and thus
\begin{equation}
 \label{caro}
 \{\psi(e_j),\psi^*(e_k)\}e_I=0.
\end{equation}
If $k<j$, then $\epsilon_{I\cup\{k\}}(j)=-\epsilon_I(j)$ by
(\ref{jkmoins}),
$\epsilon_{(I\setminus\{j\})\cup\{k\}}(k)=\epsilon_{I\cup\{k\}} (k)$
by (\ref{jkplus}) and thus (\ref{caro}) is satisfied again. 2) We
assume $j\notin I$ and $k\in I$. Then we have:
\begin{equation}
 \label{jkI}
\psi^*(e_k)e_I=0=
\psi(e_j)e_I,
\end{equation}
hence (\ref{caro}) is trivially satisfied. 3) We assume $j\in I$ and
$k\in I$. Then $k\in I\setminus\{j\}$ and so 
\begin{equation}
 \label{}
 \{\psi(e_j),\psi^*(e_k)\}e_I=\psi^*(e_k)\psi(e_j)e_I=\epsilon_I(j)\psi^*(e_k)e_{I\setminus\{j\}}=0.
\end{equation}
4) Finally we assume $j\notin I$ and
$k\notin I$. Then $\psi(e_j)e_I=0$ and $j\notin I\cup\{k\}$, hence
\begin{equation}
 \label{}
 \{\psi(e_j),\psi^*(e_k)\}e_I=\psi(e_j)\psi^*(e_k)e_I=\epsilon_{I\cup\{k\}}(k)\psi(e_j)e_{I\cup\{k\}}=0,
\end{equation}
and the proof of (\ref{car}) is complete.\\

To prove (\ref{carpsi}), we distinguish the different cases again. 1)
If $j\notin I$, $k\notin I$ then $\psi(e_j)e_I=\psi(e_k)e_I=0$ and
obviously we have
\begin{equation}
 \label{carpso}
 \{\psi(e_j),\psi(e_k)\}e_I=0.
\end{equation}
2) If $j\notin I$, $k\in I$, we have $\psi(e_j)e_I=0$ and
\begin{equation}
 \label{}
 \{\psi(e_j),\psi(e_k)\}e_I=\epsilon_I(k)\psi(e_j)e_{I\setminus\{k\}}=0
\end{equation}
since $j\notin I\setminus\{k\}$. 3) The case $j\in I$, $k\notin I$ is
analogous. 4) Now if $j\in I$, $k\in I$, $j\neq k$, we compute
\begin{equation}
 \label{kkpso}
  \{\psi(e_j),\psi(e_k)\}e_I=\left(\epsilon_I(k)\epsilon_{I\setminus\{k\}}(j)+\epsilon_I(j)\epsilon_{I\setminus\{j\}}(k)\right)e_{I\setminus\{j,k\}}.
\end{equation}
If $j<k$, then (\ref{jkplus}) implies that
$\epsilon_I(j)=\epsilon_{I\setminus\{k\}}(j)$ and (\ref{jkmoins})
assures that $\epsilon_I(k)=-\epsilon_{I\setminus\{j\}}(k)$, and we
conclude that (\ref{carpso}) follows from (\ref{kkpso}). 5) Finally if
$j=k\in I$, $\psi(e_j)\psi(e_j)e_I=\epsilon_I(j)
\psi(e_j)e_{I\setminus\{j\}}=0$. The proof of (\ref{carpsi}) is
achieved. Moreover (\ref{carstar}) follows from (\ref{carpsi}) by
taking the adjoint.\\

To end, since the quantum fields map $\delta_I$ to zero, or
$\pm\delta_{I\setminus\{j\}}$, or $\pm\delta_{I\cup\{j\}}$, they leave
invariant $l^2([A]_X)$.
\fin
We now define $\psi$ and $\psi^*$ on the whole space $l^2(X)$ by the
usual way used in the separable case \cite{dimocksea}. If $u\in l^2(X)$ is expressed as
\begin{equation}
 \label{}
 u=\sum_{k\in \NN}c_ke_{j_k},\;\;c_k\in\CC,\;\;\sum_{k\in \NN}\mid c_k\mid^2<\infty,
\end{equation}
we introduce for any $N\in\NN$
\begin{equation}
 \label{}
 \psi_N(u):=\sum_{k\leq N}c^*_k\psi(e_{j_k}),\;\; \psi_N^*(u):=\sum_{k\leq
 N}c_k\psi^*(e_{j_k})=[\psi_N(u)]^*.
\end{equation}
Given two integers $N,M$ with $M\geq N+1$, we deduce from (\ref{car})
that for any $V\in l^2(\mathcal{P}(X))$ we have
$$
<V;\{\psi_M(u)-\psi_N(u), \psi^*_M(u)-\psi^*_N(u)\}V>=\left(\sum_{k=N+1}^M\mid
c_k\mid^2\right)\Vert V\Vert^2.
$$
Therefore
$$
\Vert \psi_M(u)-\psi_N(u)\Vert^2_{\mathcal{L}(l^2(\mathcal{P}(X)))}+
\Vert \psi^*_M(u)-\psi^*_N(u)\Vert^2_{\mathcal{L}(l^2(\mathcal{P}(X)))}\leq \sum_{k=N+1}^M\mid
c_k\mid^2.
$$
We conclude that we may define the quantum fields
\begin{equation}
 \label{zepsis}
 \psi(u):=\lim_{N\rightarrow\infty}\psi_N(u),\;\;\psi^*(u):=\lim_{N\rightarrow\infty}\psi^*_N(u)\;\;in\;\;\mathcal{L}(l^2(\mathcal{P}(X))).
\end{equation}

The previous lemma directly implies the main properties of the quantum fields:
\begin{Theorem}
$\psi$ is an anti-linear map from $l^2(X)$ to
$\mathcal{L}(l^2(\mathcal{P}(X)))$ and we have for any $u,v\in l^2(X)$
\begin{equation}
 \label{karkad}
 [\psi(u)]^*=\psi^*(u),
\end{equation}
 \begin{equation}
 \label{unXX}
\Vert\psi(u)\Vert_{\mathcal{L}(l^2(\mathcal{P}(X)))}=\Vert u\Vert_{l^2(X)},
\end{equation}
 \begin{equation}
 \label{cara}
\{\psi(u),\psi^*(v)\}=<u,v>Id_{l^2(\mathcal{P}(X))},
\end{equation}
\begin{equation}
 \label{zo}
 \{\psi(u),\psi(v)\}=0,
\end{equation}
\begin{equation}
 \label{zost}
 \{\psi^*(u),\psi^*(v)\}=0,
\end{equation}
\begin{equation}
 \label{}
 \forall A\subset X,\;\forall u\in l^2(X),\;\;\psi(u)l^2([A]_X)\subset l^2([A]_X),
\end{equation}
\begin{equation}
 \label{parX}
\forall A\subset X,\; \forall u\in l^2(X),\;\;P_A(u)=0\Rightarrow\psi(u)e_A=0,
\end{equation}
\begin{equation}
 \label{trouX}
 \forall A\subset X,\; \forall u\in l^2(X),\;\;P_{X\setminus A}(u)=0\Rightarrow\psi^*(u)e_A=0.
\end{equation}

 \label{zeteoq}
\end{Theorem}


The quantum field provided by the previous theorem is not unique since
it depends on the choices of the
linear order on $X$, and  the function
$\epsilon$. Nevertheless, the representation of the CAR on $l^2(X)$
given by $(\psi,l^2([A]_X))$
is actually
unique up to an unitary transform and it depends only on $A$
and $X\setminus A$. We denote
\begin{equation}
 \label{}
\aleph_-:=\mid A\mid,\;\;\aleph_+:=\mid X\setminus A\mid.
\end{equation}
We take a bijection $\theta_-$ from $A$ onto $\aleph_-$
and a bijection $\theta_+$ from $X\setminus A$ onto $\aleph_+$.
We denote $\theta$ the bijection from $X$ onto
$\aleph_-\sqcup\aleph_+$ defined by
\begin{equation}
 \label{}
 j\in A\Rightarrow \theta(j):=\theta_-(j),\;\; j\in X\setminus A\Rightarrow \theta(j):=\theta_+(j).
\end{equation}
Therefore the map
\begin{equation}
 \label{}
 u\in l^2(X)\longmapsto u\circ\theta^{-1}\in l^2(\aleph_-\sqcup\aleph_+)=l^2(\aleph_-)\oplus l^2(\aleph_+)
\end{equation}
is an isometry.

We consider an anti-unitary operator $\mathrm C$ on $l^2(\aleph_-)\oplus
l^2(\aleph_+)$ and we introduce the fermionic
Fock space
\begin{equation}
 \label{}
 \mathfrak{H}_{\mathcal{F}}:=\mathcal{F}^{\wedge}(\mathrm{C}l^2(\aleph_-))\otimes \mathcal{F}^{\wedge}(l^2(\aleph_+)).
\end{equation}
The Fock quantization 
$(\mathfrak{H}_{\mathcal{F}},\Psi_{\mathcal{F}})$ on
$l^2(\aleph_-)\oplus l^2(\aleph_+)$ is defined by choosing the
antilinear map, $\Psi_{\mathcal F}$,  from
$l^2(\aleph_-)\oplus l^2(\aleph_+)$ to
$\mathcal{L}(\mathfrak{H}_{\mathcal F})$ defined by 
\begin{equation}
 \label{}
 \Psi_{\mathcal{F}}(u_-\oplus u_+):=a_-^*(\mathrm{C}u_-)\otimes (-1)^{\mathrm N_+}+Id\otimes
 a_+(u_+),\;\;u_{\pm}\in l^2(\aleph_{\pm}),
\end{equation}
where $a_-^*$ (resp. $a_+$ ) is the creation (resp. annihilation)
operator on $\mathcal{F}^{\wedge}\left(\mathrm{C}l^2(\aleph_{-})\right)$ (resp. $\mathcal{F}^{\wedge}\left(l^2(\aleph_+)\right)$), and $\mathrm{N}_+$
is the number operator on $\mathcal{F}^{\wedge}(l^2(\aleph_+))$. The {\it Fock
  vacuum} is the vector
\begin{equation}
 \label{}
 \Omega_{\mathcal{F}}:=(1,0,0,...)\in[\mathrm{C}l^2(\aleph_-)]^{\wedge 0}\otimes [l^2(\aleph_-)]^{\wedge 0}
\end{equation}
that obviously satisfies
\begin{equation}
 \label{}
 \forall u_{\pm}\in l^2(\aleph_{\pm}),\;\;\Psi_{\mathcal{F}}(u_+)\Omega_{\mathcal F}=0,\;\; [\Psi_{\mathcal{F}}(u_-)]^*\Omega_{\mathcal F}=0.
\end{equation}
We know that this representation is irreducible.

\begin{Theorem}
For any $A\subset X$, the quantization $(\psi,l^2([A]_X)$ is an
irreducible representation of the CAR on $l^2(X)$, moreover there
exists a unitary transformation $\UU_A$ from $\mathfrak{H}_{\mathcal F}$ onto $
l^2([A]_X)$ such that for any $u\in l^2(X)$ we have
\begin{equation}
 \label{uniteq}
\psi(u) =\UU_A \Psi_{\mathcal{F}}\left(u\circ\theta^{-1}\right)\UU_A^{-1}\;\;on\;\;l^2([A]_X).
\end{equation}
 \label{theof}
\end{Theorem}

{\it Proof.} The irreducibility of the representation
$(\psi,l^2([A]_X)$ follows from (\ref{uniteq}) since the Fock
representation is irreducible. We  express any subset $I_F\in \mathcal{P}_F(X)$ as a
finite decreasing sequence
$$
I_F=\{j_n\;\;1\leq n\leq N,\;\;j_N<j_{N-1}<...<j_1\},
$$
and if $\varphi=\psi, \;\psi^*$ we denote
$$
\prod_{j\in I_F}\varphi(e_j):=\varphi(e_{j_N})\varphi(e_{j_{N-1}})...\varphi(e_{j_1}),
$$
and for $\varphi=\Psi_{\mathcal F},\; \Psi^*_{\mathcal F}$ we denote
$$
\prod_{j\in I_F}\varphi(e_{\theta(j)}):=\varphi(e_{\theta(j_N)})\varphi(e_{\theta(j_{N-1})})...\varphi(e_{\theta(j_1)}).
$$
We note that (\ref{decomposI}) assures that
 any $I\in[A]_X$ can be uniquely written as $I=(A\setminus I_-)\sqcup I_+$,
 and we have
$$
\prod_{j\in I_+}\psi^*(e_j)\prod_{k\in I_-}\psi(e_k)e_A=\varepsilon e_I,\;\;\varepsilon\in\{-1,1\}.
$$
We conclude that the set
\begin{equation}
 \label{basisI}
 \left\{\prod_{j\in I_+}\psi^*(e_j)\prod_{k\in I_-}\psi(e_k)e_A;\;\;I_-\in{\mathcal{P}_F(A)},\;\;I_+\in\mathcal{P}_F(X\setminus
   A)\right\},
\end{equation}
is a Hilbert basis of $l^2([A]_X)$.

Since
$\{e_{\theta_-(j)};\;\;j\in A\}$ is a Hilbert basis of $l^2(\aleph_-)$
and $\{e_{\theta_+(j)};\;\;j\in X\setminus A\}$ is a Hilbert basis of
$l^2(\aleph_+)$, we know that
\begin{equation}
 \label{}
 \left\{\prod_{j\in
     I_+}\Psi_{\mathcal{F}}^*(e_{\theta_+(j)})\prod_{k\in I_-}\Psi_{\mathcal{F}}(e_{\theta_-(k)})\Omega_{\mathcal{F}};\;\;I_-\in{\mathcal{P}_F(A)},\;\;I_+\in\mathcal{P}_F(X\setminus
   A)\right\}
\end{equation}
is a Hilbert basis of $\mathfrak{H}_{\mathcal{F}}$. We define the unitary transformation $\UU_A$ from $\mathfrak{H}_{\mathcal{F}}$ onto $
l^2([A]_X)$ by putting
\begin{equation}
 \label{}
 \UU_A\left(\prod_{j\in
     I_+}\Psi_{\mathcal{F}}^*(e_{\theta_+(j)})\prod_{k\in I_-}\Psi_{\mathcal{F}}(e_{\theta_-(k)})\Omega_{\mathcal{F}}\right)=\prod_{j\in I_+}\psi^*(e_j)\prod_{k\in I_-}\psi(e_k)e_A.
\end{equation}
Now to prove (\ref{uniteq}) it is sufficient to show that for anay
$l\in X$, $I_-\in\mathcal{P}_F(A)$, $I_+\in\mathcal{P}_F(X\setminus
A)$ we have
\begin{equation}
 \label{egalop}
 \psi(e_l) \prod_{j\in I_+}\psi^*(e_j)\prod_{k\in I_-}\psi(e_k)e_A=\UU_A\Psi_{\mathcal{F}}\left(e_{\theta(l)}\right) \prod_{j\in
     I_+}\Psi_{\mathcal{F}}^*(e_{\theta_+(j)})\prod_{k\in I_-}\Psi_{\mathcal{F}}(e_{\theta_-(k)})\Omega_{\mathcal{F}}.
 \end{equation}
 The CAR assure that:\\
 
 1) If $l\notin I_+$  there exists
 $\varepsilon\in\{-1,+1\}$ such that
 $$
\psi(e_l) \prod_{j\in I_+}\psi^*(e_j)\prod_{k\in
  I_-}\psi(e_k)e_A=
\left\{
  \begin{array}{c}
    0\;\;if\;\;l\notin A,\\
    \varepsilon\prod_{j\in I_+}\psi^*(e_j)\prod_{k\in
  I_-\cup\{l\}}\psi(e_k)e_A\;\;if \;\;l\in A,
  \end{array}
  \right.
  $$
  $$
\Psi_{\mathcal{F}}\left(e_{\theta(l)}\right) \prod_{j\in
     I_+}\Psi_{\mathcal{F}}^*(e_{\theta_+(j)})\prod_{k\in
     I_-}\Psi_{\mathcal{F}}(e_{\theta_-(k)})\Omega_{\mathcal{F}}=
   \left\{
  \begin{array}{c}
    0\;\;if\;\;l\notin A,\\
    \varepsilon \prod_{j\in
     I_+}\Psi_{\mathcal{F}}^*(e_{\theta_+(j)})\prod_{k\in
     I_-\cup\{l\}}\Psi_{\mathcal{F}}(e_{\theta_-(k)})\Omega_{\mathcal{F}}\;\;if \;\;l\in A,
  \end{array}
  \right.
  $$
  and we conclude that (\ref{egalop}) is satisfied;\\

  2) If $l\in I_+$ there exists $\varepsilon\in\{-1,+1\}$ such that
  $$
\psi(e_l) \prod_{j\in I_+}\psi^*(e_j)\prod_{k\in
  I_-}\psi(e_k)e_A=\varepsilon\prod_{j\in I_+\setminus\{l\}}\psi^*(e_j)\prod_{k\in
  I_-}\psi(e_k)e_A,
$$
$$
\Psi_{\mathcal{F}}\left(e_{\theta(l)}\right) \prod_{j\in
     I_+}\Psi_{\mathcal{F}}^*(e_{\theta_+(j)})\prod_{k\in
     I_-}\Psi_{\mathcal{F}}(e_{\theta_-(k)})\Omega_{\mathcal{F}}=\varepsilon\prod_{j\in
     I_+\setminus\{l\}}\Psi_{\mathcal{F}}^*(e_{\theta_+(j)})\prod_{k\in
     I_-}\Psi_{\mathcal{F}}(e_{\theta_-(k)})\Omega_{\mathcal{F}},
   $$
   hence  (\ref{egalop}) is satisfied again. The proof is complete.
   \fin
We deduce a form of the Shale-Stinespring criterion of implementation of
unitary operators, adapted to our quantization.
   \begin{Corollary}
Given $A\subset X$, let $U$ be an unitary operator on $l^2(X)=l^2(A)\oplus l^2(X\setminus
A)$. Then there
exists a unitary operator $\UU$ on $l^2([A]_X)$ satisfying for any
$u\in l^2(X)$
\begin{equation}
 \label{implu}
 \psi(Uu)=\UU\psi(u)\UU^{-1}
\end{equation}
if and only if  $P_AUP_{X\setminus A}$ and $P_{X\setminus A}UP_A$ are
Hilbert-Schmidt operators. 
     \label{shal}
   \end{Corollary}

   {\it Proof.}
   With the previous notations we have
   $$
\psi(Uu)=\UU_A\Psi_{\mathcal{F}}((Uu)\circ\theta^{-1})\UU_A^{-1}=\UU_A\Psi_{\mathcal{F}}(\mathcal
U(u\circ\theta^{-1}))\UU_A^{-1}
$$
where $\mathcal U$ is the unitary operator on
$l^2(\aleph_-\sqcup\aleph_+)$ defined by
   $$
v\in l^2(\aleph_-\sqcup\aleph_+)\longmapsto\mathcal{U}v:=[U(v\circ\theta)]\circ\theta^{-1}.
   $$
Therefore $\UU$ exists and satisfies (\ref{implu}) iff $\mathcal{U}$ is
implementable in the Fock representation of
$l^2(\aleph_-)\oplus l^2(\aleph_+)$. The famous theorem of Shale and
Stinespring states that a necessary and sufficient condition is that
$P_{\aleph_-}\mathcal{U}P_{\aleph_+}$ and
$P_{\aleph_+}\mathcal{U}P_{\aleph_-}$ are Hilbert-Schmidt
operators. We have
   $$
   P_{\aleph_-}\mathcal{U}P_{\aleph_+}v=[P_AUP_{X\setminus A}(v\circ\theta)]\circ\theta^{-1}
   $$
   hence we deduce that
   \begin{equation}
     \label{}
     \begin{split}
       \sum_{y,y'\in\aleph_-\sqcup\aleph_+}\mid<P_{\aleph_-}\mathcal{U}P_{\aleph_+}e_y;e_{y'}>\mid^2&=\sum_{y_-\in\aleph_-}\sum_{y_+\in\aleph_+}\mid<\mathcal{U}e_{y_+};e_{y_-}\mid^2\\
       &=\sum_{j_-\in A}\sum_{j_+\in X\setminus
         A}\mid<Ue_{j_+};e_{j_-}\mid^2\\
       &=\sum_{j,j'\in X}\mid<P_{A}UP_{X\setminus A}e_y;e_{y'}>\mid^2.
       \end{split}
 \end{equation}
  We conclude that $P_{\aleph_-}\mathcal{U}P_{\aleph_+}$ is
  Hilbert-Schmidt on $l^2(\aleph_-\sqcup\aleph_+)$ iff
  $P_{A}UP_{X\setminus A}$ is Hilbert-Schmidt on $l^2(X)$. And the
  same holds for $P_{\aleph_+}\mathcal{U}P_{\aleph_-}$ and
    $P_{X\setminus A}UP_{A}$. The proof is complete.
   \fin

Finally we prove that, up to a unitary transform, our irreducible
representations depend only on $\mid A\mid$ and $\mid X\setminus A\mid$.
We consider two totally ordered infinite sets $X_i$, $i=1,2$ and two
  epsilon functions $\epsilon_i$ given by Corollary \ref{}, and the
  quantum fields $\psi_i$ defined on $l^2(X_i)$ as previously. Let
  $A_i$ be in $\mathcal{P}(X_i)$ such that
  \begin{equation}
 \label{}
 \mid A_1\mid=\mid A_2\mid,\;\;  \mid X_1\setminus A_1\mid=\mid
 X_2\setminus A_2\mid,
\end{equation}
and we choose a bijection $\Theta$ from $X_1$ onto $X_2$ satisfying
\begin{equation}
 \label{}
 \Theta(A_1)=A_2.
\end{equation}

\begin{Theorem}
 There exists a unitary transform $\UU$ from $l^2([A_2]_{X_2})$ onto
 $l^2([A_1]_{X_1})$ such that for any $u_2\in l^2(X_2)$ we have
 \begin{equation}
 \label{}
 \psi_2(u_2)=\UU^{-1}\psi_1(u_2\circ\Theta)\UU\;\;on\;\;l^2([A_2]_{X_2}).
\end{equation}
 \label{theoequiv}
\end{Theorem}

{\it Proof.}
We denote
$\aleph_-:=\mid A_i\mid$, $\aleph_+:=\mid
X_i\setminus A_i\mid$. We introduce a bijection $\theta_i$ from $X_i$
onto $\aleph_-\sqcup\aleph_+$ such that $\theta_i(A_i)=\aleph_-$,
$\theta_i(X_i\setminus A_i)=\aleph_+$. The previous Theorem assures there
exists a unitary transform $U_i$ from
$\mathcal{F}^{\wedge}(\mathrm{C}l^2(\aleph_-))\otimes
\mathcal{F}^{\wedge}(l^2(\aleph_+))$ onto $l^2([A_i]_{X_i})$ such that
for any $u_i\in l^2(X_i)$ we have
\begin{equation}
 \label{}
 \psi_i(u_i)=U_i\Psi_{\mathcal F}\left(u_i\circ \theta_i^{-1}\right)U_i^{-1}.
\end{equation}
Now we introduce an unitary operator $T$ on $l^2(\aleph_-)\oplus
l^2(\aleph_+)$ by putting
\begin{equation}
 \label{}
 T:\;v\in l^2(\aleph_-)\oplus
l^2(\aleph_+)\longmapsto v\circ \theta_2\circ \Theta\circ \theta_1^{-1}\in l^2(\aleph_-)\oplus
l^2(\aleph_+).
\end{equation}
Since $T$ leaves invariant $l^2(\aleph_{\pm})$ it is implementable in
the Fock representation: there exists a unitary operator $\TT$ on $\mathcal{F}^{\wedge}(\mathrm{C}l^2(\aleph_-))\otimes
\mathcal{F}^{\wedge}(l^2(\aleph_+))$ such
that for any $v\in l^2(\aleph_-)\oplus l^2(\aleph_+)$ we have
$$
\Psi_{\mathcal F}(Tv)=\TT \Psi_{\mathcal F}(v)\TT^{-1}.
$$
Since the previous Theorem assures that there exist unitary
tranformations $\UU_{A_i}$ from $\mathfrak{H}_{\mathcal{F}}$ onto $
l^2([A_i]_{X_i})$  such that for any $u_i\in l^2(X_i)$
$$
\psi_i(u_i) =\UU_{A_i} \Psi_{\mathcal{F}}\left(u_i\circ\theta_i^{-1}\right)\UU_{A_i}^{-1}\;\;on\;\;l^2([A_i]_{X_i}),
$$
we compute
\begin{equation}
 \label{}
 \begin{split}
   \psi_1(u_2\circ\Theta)&=\UU_1\Psi_{\mathcal
     F}\left(u_2\circ\theta_2^{-1}\circ(\theta_2\circ\Theta\circ\theta_1^{-1})\right)\UU_1^{-1}\\
   &=\UU_1\TT\Psi_{\mathcal
     F}\left(u_2\circ\theta_2^{-1}\right)\TT^{-1}\UU_1^{-1}\\
   &=\UU_1\TT\UU_2^{-1}\psi_2(u_2)\UU_2\TT\UU_1^{-1}.
   \end{split}
 \end{equation}
 Now it is sufficient to choose
 $$
\UU:=\UU_1\TT\UU_2^{-1}.
 $$
 \fin

 Taking $X_1=X_2$, $A_1=A_2$, $\Theta=Id$, we immediately deduce the following
 \begin{Corollary}
 Given an infinite set $X$, let $\psi_1$, $\psi_2$ be the quantizations associated to two linear
 orders on $X$ and two $\epsilon$-functions. Then for any $A\subset X$
 there exists a unitary operator $\UU$ on $l^2([A]_{X})$ such that for any $u\in l^2(X)$ we have
 \begin{equation}
 \label{}
 \psi_2(u)=\UU^{-1}\psi_1(u)\UU\;\;on\;\;l^2([A]_{X}).
\end{equation}
 \label{corox}
\end{Corollary}
\section{Application to the Dirac theory}

We consider  two Hilbert spaces $\mathfrak{h}_{\pm}$, possibly non-separable, and we define
\begin{equation}
 \label{}
 \mathfrak{h}=\mathfrak{h}_-\oplus\mathfrak{h}_+,
\end{equation}and we denote $P_{\pm}$ the orthogonal projector on $\mathfrak{h}_{\pm}$.
Invoking the Axiom of Choice, we take two Hilbert basis $(\mathfrak{e}_j)_{j\in
  X_{\pm}}$ of $\mathfrak{h}_{\pm}$. We put
\begin{equation}
 \label{}
 X:=X_-\sqcup X_+,
\end{equation}
then $(\mathfrak{e}_j)_{j\in X}$ is a Hilbert basis of $\mathfrak{h}$, and using the Axiom of Choice again, we endow $X$ with a total order.
Finally, we use a last time this Axiom to get an ultrafilter on
$\mathcal{P}_F(X)\setminus\{\emptyset\}$ to
obtain an $\epsilon$-function given by Corollary \ref{corolpa}. $\psi$
being the quantum field on $l^2(X)$ given by Theorem \ref{zeteoq},
$(\psi,l^2([X_-]_X)$ is an irreducible representation of the CAR on
$l^2(X)$. Now we introduce the canonical isometry $\mathcal{J}$ from $\mathfrak{h}$
onto $l^2(X)$,
\begin{equation}
 \label{}
 \mathfrak{u}\in\mathfrak{h}\longmapsto\mathcal{J}(\mathfrak{u})=\left(<\mathfrak{e}_j;\mathfrak{u}>\right)_{j\in
   X}\in l^2(X),
\end{equation}
and we define a quantum field on $\mathfrak{h}$ by putting
\begin{equation}
 \label{zeqq}
 \Psi_D(\mathfrak{u}):=\psi\left(\mathcal{J}(\mathfrak{u})\right)\in\mathcal{L}(\mathfrak{H}_D),\;\;\mathfrak{H}_D:=l^2([X_-]_X).
\end{equation}
The {\it Dirac sea} describing the state
fulfilled by $\mathfrak{h}_-$ is the vector
\begin{equation}
 \label{}
 \Omega_D:=e_{X_-}\in \mathfrak{H}_D.
\end{equation}

 These mathematical objects have the usual meaning of the Dirac theory: $\mathfrak{h}_{+(-)}$ is the space of the
classical solutions of positive (negative) energy. The quantum field $\Psi_D$ is an operator of
annihilation: $\Psi_D(\mathfrak{e}_j)$ annihilates the state
$\mathfrak{e}_j$, therefore it annihilates a particle in the state
$\mathfrak{e}_j$ when $j\in X_+$, and it creates a hole in the Dirac
sea when $j\in X_-$.

We shall compare with the classic Fock
quantization defined by the Hilbert space
\begin{equation}
 \label{}
 \mathfrak{H}_{\mathcal F}:=\mathcal{F}^{\wedge}\left(\mathcal{C}\mathfrak{h}_{-}\right)\otimes\mathcal{F}^{\wedge}\left(\mathfrak{h}_+\right),
\end{equation}
the Fock vacuum vector
\begin{equation}
 \label{}
 \Omega_{\mathcal F}:=(1,0,0,0...)\in
 \left[\mathcal{C}\mathfrak{h}_{-}\right]^{\wedge 0}\otimes \left[\mathfrak{h}_{+}\right]^{\wedge 0},
\end{equation}
and the quantum field
\begin{equation}
 \label{}
 \Psi_{\mathcal{F}}(u):=a_-^*(\mathcal{C}P_-u)\otimes (-1)^{\mathrm N_+}+Id\otimes
 a_+(P_+u)\in\mathcal{L}\left( \mathfrak{H}_{\mathcal{F}}\right),
\end{equation}
where $a_-^*$ (resp. $a_+$ ) is the creation (resp. annihilation)
operator on $\mathcal{F}^{\wedge}\left(\mathcal{C}\mathfrak{h}_{-}\right)$ (resp. $\mathcal{F}^{\wedge}\left(\mathfrak{h}_+\right)$), $\mathrm{N}_+$
is the number operator on $\mathcal{F}^{\wedge}(\mathfrak{h}_+)$ and
$\mathcal{C}$ is an anti-unitary operator on $\mathfrak{h}$. Now
$\mathfrak{h}_+$ is the one-particle space, $\mathcal{C}\mathfrak{h}_-$
is the one-antiparticle space, and
$\Psi_{\mathcal{F}}(\mathfrak{e}_j)$ annihilates a particle if $j\in
X_+$ and creates an antiparticle if $j\in X_-$.

\begin{Theorem}
 $(\Psi_D,\mathfrak{H}_D)$  defined by (\ref{zeqq}) is an irreducible representation of the CAR on
$\mathfrak{h}$. All the representations obtained by changing  the
total order on $X$, or the $\epsilon$-function, or the
Hilbert basis, are unitarily equivalent. Moreover there exists a
unitary transform $\UU$ from $\mathfrak{H}_{\mathcal F}$ onto $\mathfrak{H}_D$ such
that
\begin{equation}
 \label{}
 \Omega_D=\UU\Omega_{\mathcal F},
\end{equation}
\begin{equation}
 \label{}
 \forall u\in\mathfrak{h},\;\;\Psi_D(u)=\UU\Psi_{\mathcal F}(u)\UU^{-1}.
\end{equation}
 \label{teodulis}
\end{Theorem}

{\it Proof.}
If we change the total order on $X$ or the $\epsilon$-function,
Corollary \ref{corox} shows that we obtain a new representation that is
unitarily equivalent to the previous one. More generally, if we take
another basis $(\mathfrak{e}'_j)_{j\in X'_{\pm}}$ of $\mathfrak{h}_{\pm}$,
  Theorem \ref{theoequiv} assures that the quantizations built from
  these two choices of basis are unitarily equivalent. Moreover, if we
  take an anti-unitary transform $\mathcal{C}$ on $\mathfrak{h}$,
  Theorem \ref{theof} implies that these quantizations are unitarily
  equivalent to the Fock quantization $\Psi_{\mathcal F}$ on $l^2(\aleph_-)\oplus
  l^2(\aleph_+)$ where $X_{\pm}=\aleph_{\pm}=dim(\mathfrak{h}_{\pm})$ and $\mathrm{C}=\mathcal{J}\mathcal{C}\mathcal{J}^{-1}$, which
  is obviously unitarily equivalent to the classic Fock quantization
  on $\mathfrak{h}_-\oplus\mathfrak{h}_+$.

\fin

We introduce a general setting for the abstract Dirac equations that
have the form
\begin{equation}
 \label{}
 i\frac{d\psi}{dt}=H\psi,
\end{equation}
with the following assumptions:

1) $H$ is a densely defined selfadjoint operator on $\mathfrak{h}$ with domain
$\mathfrak{D}(H)$;

2) there exist two densely
defined selfadjoint operators $H_{\pm}$ on $\mathfrak{h}_{\pm}$ with domain
$\mathfrak{D}(H_{\pm})$ such that
\begin{equation}
 \label{}
 0\leq H_+,
 \end{equation}
 \begin{equation}
 \label{}
 H=H_-\oplus H_+,\;\;\mathfrak{D}(H)=\mathfrak{D}(H_-)\oplus \mathfrak{D}(H_+);
\end{equation}

3) There exists an anti-unitary operator $\mathcal C$ on
$\mathfrak{h}$ such that,
\begin{equation}
 \label{}
 \mathcal{C} \mathfrak{h}_{\pm}=\mathfrak{h}_{\mp},\;\;\mathcal{C} P_{\pm}=P_{\mp}\mathcal{C},
\end{equation}
\begin{equation}
 \label{}
 \mathcal{C}\mathfrak{D}(H_{\pm})=\mathfrak{D}(H_{\mp}),\;\;\mathcal{C}H=-H\mathcal{C}.
\end{equation}
In fact these assumptions imply that
\begin{equation}
 \label{}
 H_-\leq 0
\end{equation}
since we have:
\begin{equation*}
 \label{}
 \begin{split}
   0&\leq <P_+u;H_+P_+u>=<\mathcal{C}H_+P_+u;\mathcal{C}P_+u>\\
   &=<\mathcal{C}P_+Hu;P_-\mathcal{C}u>=<P_-\mathcal{C}Hu;P_-\mathcal{C}u>\\
   &=-<P_-H\mathcal{C}u;P_-\mathcal{C}u>=-<H_-P_-\mathcal{C}u;P_-\mathcal{C}u>.
   \end{split}
\end{equation*}
The time evolution of the classical Dirac fields is given by the
unitary group $e^{-itH}=e^{-itH_-}\oplus e^{-itH_+}$ on $\mathfrak{h}$, that leaves invariant $\mathfrak{h}_{\pm}$.The time evolution of the quantum field is defined for $t\in\RR$,
$u\in\mathfrak{h}$ by putting:
\begin{equation}
 \label{}
 \Psi_D(t,u):=\Psi_D\left(e^{itH}u\right).
\end{equation}

\begin{Theorem}
 For any $t\in\RR$, $(\Psi_D(t,.),\mathfrak{H}_D)$ is an irreductible
 representation of the CAR on $\mathfrak{h}$ and for any $u\in\mathfrak{D}(H)$ the map
 $t\mapsto \Psi_D(t,u)\in\mathcal{L}(\mathfrak{H}_D)$ is a strongly
 differentiable function satisfying
 \begin{equation}
 \label{eqqq}
 i\frac{d}{dt}\Psi_D(t,u)=\Psi_D(t,Hu),\;\;\Psi_D(0,u)=\Psi_D(u).
\end{equation}
Moreover there exists a densely defined self-adjoint operator $\HH_D$
on $\mathfrak{H}_D$ such that
\begin{equation}
 \label{}
 0\leq\HH_D,
\end{equation}
\begin{equation}
 \label{}
 \forall t\in\RR,\;\forall u\in\mathfrak{h},\;\;\Psi_D(t,u)=e^{it\HH_D}\Psi_D(u)e^{-it\HH_D}.
\end{equation}
 \label{}
\end{Theorem}

{\it Proof.}
The first assertion is obvious since $e^{itH}$ is unitary. Moreover if $u\in\mathfrak{D}(H)$, the map $t\mapsto e^{itH}u$ belongs to
$C^1(\RR_t;\mathfrak{h})\cap C^0(\RR_t;\mathfrak{D}(H))$, therefore, since $u\mapsto\Psi_D(u)$ is an
antilinear bounded map,  we can differentiate
$\Psi_D(t,e^{itH}u)$ and we obtain (\ref{eqqq})).
Now since $\mathcal{C}\mathfrak{h}_-=\mathfrak{h}_+$, we consider
$\mathfrak{H}_{\mathcal F}=\mathcal{F}^{\wedge}(\mathfrak{h}_+)\otimes
\mathcal{F}^{\wedge}(\mathfrak{h}_+)$, and the positive  densely defined
self-adjoint operator on $\mathfrak{H}_{\mathcal F}$ 
\begin{equation}
 \label{}
 \HH_{\mathcal F}:=d\Gamma(H_+)\otimes Id+Id\otimes d\Gamma(H_+),
\end{equation}
where $\Gamma$ is the usual second-quantization functor. Therefore,
using unitary operator $\UU$ of Theorem \ref{teodulis}, the operator
\begin{equation}
 \label{}
 \HH_D:=\UU\HH_{\mathcal F}\UU^{-1}
\end{equation}
is positive, densely defined and self-adjoint on $\mathfrak{H}_D$. We have:
\begin{equation}
 \label{}
 e^{it\HH_{\mathcal F}}=\Gamma\left(e^{itH_+}\right)\otimes \Gamma\left(e^{itH_+}\right),
\end{equation}
\begin{equation}
 \label{}
 e^{it\HH_D}=\UU e^{it\HH_{\mathcal F}}\UU^{-1}.
\end{equation}
We achieve the proof by writting
\begin{equation}
 \label{}
 \begin{split}
   \Psi_D(e^{itH}u)&=\UU\Psi_\mathcal{F}(e^{itH}u)\UU^{-1}\\
   &=\UU\left( a_-^*(\mathcal{C}P_-e^{itH}u)\otimes
     (-1)^{\mathrm{N}_+}+Id\otimes a_+(P_+e^{itH}u)\right)\UU^{-1}\\
   &=\UU\left(a_-^*(e^{itH_+}P_+\mathcal{C}u)\otimes(-1)^{\mathrm{N}_+}+Id\otimes
     a_+(e^{itH_+}P_+u)\right)\UU^{-1}\\
   &=\UU e^{it\HH_{\mathcal
       F}}\Psi_{\mathcal{F}}(u)e^{-it\HH_{\mathcal F}}\UU^{-1}\\
   &=e^{it\HH_D}\Psi_D(u)e^{-it\HH_D}.
   \end{split}
 \end{equation}
 \fin


 

\end{document}